\documentclass[preprint2]{aastex}

\shorttitle{Wavefront Sensing with Donut}
\shortauthors{Tokovinin and Heathcote}

\begin{document}

\title{Donut: measuring optical aberrations from a single extra-focal image}

\author{A.~Tokovinin and S.~Heathcote}
\affil{Cerro Tololo Inter-American Observatory, Casilla 603, La Serena, Chile}
\email{atokovinin@ctio.noao.edu, sheathcote@ctio.noao.edu}

\date{\center Accepted for publication in PASP June 14, 2006}

\begin{abstract}
We propose  a practical method  to calculate Zernike  aberrations from
analysis  of  a  single  long-exposure defocused  stellar  image.   It
consists  in  fitting  the  aberration coefficients  and  seeing  blur
directly to  a  realistic  image  binned  into  detector
pixels. This "donut"  method is different from curvature  sensing in that
it does not make the usual  approximation of  linearity.  We calculate  the
sensitivity  of  this  technique  to  detector and  photon  noise  and
determine optimal  parameters for some  representative cases. Aliasing
of high-order un-modeled  aberrations is evaluated and  shown to be
similar  to a  low-order Shack-Hartmann  sensor. The  method  has been
tested  with  real data  from  the   SOAR and Blanco 4m telescopes. \\  
%
\end{abstract}

\keywords{telescopes}

\section{Introduction}

An experienced optician can detect low-order aberrations by looking at
the defocused  image of a  point source, and  it is trivial  to obtain
defocused images  with modern telescopes equipped  with CCD detectors.
Yet, measurements  of low-order  aberrations including focus  are still
made  by  indirect techniques,  or  using  special  equipment such  as
Shack-Hartmann (S-H)  sensors.  Astronomers spend  significant time in
acquiring  ``focus sequences''  of  stellar images,  then fitting  the
image  half-width  vs.  focus  curve  with  a  parabola  to  find  the
best-focus position.

The appeal of estimating aberrations directly from defocused images is
evident.   No  special equipment  is  needed  apart  from a  regular
imager.   The  aberrations in  the  true  science  beam are  measured,
including all optics of  the instrument but excluding additional optics
of a wave-front  sensor.  The amount of defocus  is easily adjustable,
providing flexibility.

It  was  recognized  since long  time  that optical  aberrations  cannot  be
retrieved   from  a   focused  image   of  a   point   source  without
ambiguity. However, combining {\em two} images with a known difference
of aberration provides a solution  to this problem, even for non-point
sources.  The method of {\it phase diversity} which exploits this idea
has been used since the beginning of the 80-s \citep{Thelen99}.  Phase
diversity  works well  when the  image is  sampled to  the diffraction
limit, e.g. in adaptive optics  \citep{Hartung}.  This is not the case
for conventional astronomical imagery with a pixel size matched to the
seeing.    Yet  another   method  for   extracting   aberrations  from
well-sampled focused images  by means of a trained  neural network was
suggested by Sandler and later tried by \citet{LH92}. The authors note
that their method is  extremely computationally intensive and has some
subtleties. To our knowledge, this method is not in use nowadays.

The relation of the intensity distribution in a defocused image to the
local  wavefront curvature  is described  by the  so-called irradiance
transport equation  \citep{Roddier90}. This relation is  basic to {\it
curvature sensing}  as used  in adaptive optics  \citep{Roddier99}.  A
commercial software package for telescope aberration analysis based on
the   same  principle  has   been  developed   by  Northcott\footnote{
Northcott,  M.J., The  {\it  ef} wavefront  reduction package.   1996,
Laplacian  Optics  Inc.}  and  is  used  at  some observatories.  This
method, however, is not very  practical because it requires two images
with relatively large and equal defocus of opposite sign.

The need  of two images for  curvature sensing has  been questioned by
\citet{Hickson94}  who shows  that  even in  the  context of  adaptive
optics a single extra-focal image  is sufficient and provides a better
signal-to-noise  ratio with  a  CCD detector  and  faint guide  stars,
despite   scintillation   noise.    One   image  is   sufficient   for
non-ambiguous  aberration retrieval  as long  as the  rays originating
from different  parts of  the aperture do  not cross each  other, i.e.
for a  sufficiently large defocus  that avoids caustics.   The minimum
defocus is proportional to  the amplitude of higher-order aberrations.
\citet{Ragazzoni} have used this technique in their experiment.

The intensity  transport equation  is not valid  for a  small defocus,
where physical  optics must be used  instead. However, this  is not an
obstacle for sensing low-order aberrations,  as long as they are small
enough,  so that  a relation  between aberration  and  image intensity
remains  linear.   \citet{Bharmal}  develop  such  near-focus  sensing
technique  for low-order  adaptive  optics, providing  in their  paper
several valuable  insights into  this problem.  However,  their method
still requires two images, intra- and extra-focal.

Here we present a quantitative method of measuring optical aberrations
from  a single  defocused image.   Such images  often  resemble donuts
(because of the shadow at the center caused by the central obscuration
in a Cassegrain telescope), so we call this technique ``donut''.  This
work  is primarily  motivated  by  the need  for  a simple  wave-front
sensing     method    for    the     SOAR    telescope     in    Chile
\citep{Sebring98,Krab04}.  All numerical  examples in the article were
computed for a telescope diameter $D=4.1$~m with a central obscuration
0.24,  appropriate  for SOAR.   The  proposed  technique is  primarily
intended for  active optics, it  is too slow for  real-time 
correction of turbulence.

The donut method is different  from standard curvature sensing. We use
physical optics and directly fit a model of the aberrated image to the
measured  ``donut''. The  initial approximation  is obtained  from the
second  moments   of  the  intensity  distribution   as  described  in
Sect.~\ref{sec:mom}.  Then an iterative fitting algorithm presented in
Sect.~\ref{sec:fit}, with further details  in the Appendix, is used to
refine   the   model   including   higher   order   aberrations.    In
Sect.~\ref{sec:perf} we evaluate the errors of aberrations measured by
this method and compare it  to a low-order Shack-Hartmann sensor while
  examples of  actual  performance are  given in  Sect.~\ref{sec:exa}.
Finally we present our conclusions in Sect.~\ref{sec:concl}.

\section{Image formation}
\label{sec:image}

To  begin the  presentation of  our algorithm  we recall  the textbook
theory  of image  formation, e.g.   \citep{BW}. Let  ${\bf a}$  be the
2-dimensional angular  coordinate in the image plane  (in radians) and
${\bf x}$ -- the coordinate in the plane of telescope pupil. The shape
of the wave-front  is $W({\bf x})$ and the phase of  the light wave is
$\phi({\bf  x})  = (2  \pi/\lambda)  W({\bf  x})$  for the  wavelength
$\lambda$. Then the intensity  distribution in the image plane $I({\bf
a})$ is computed as

\begin{equation}
I({\bf a}) = I_0 \left| 
\int P ({\bf x}) e ^{ i \phi({\bf x}) - 2 \pi i {\bf x}  {\bf
    a}/\lambda } \; {\rm d}^2 {\bf
  x} \right| ^2 ,
\label{eq:I}
\end{equation}
where  $  P({\bf x})$  is  the  pupil  transmission function  and  the
normalization constant $I_0$ is of no importance here.

\begin{figure}
\plotone{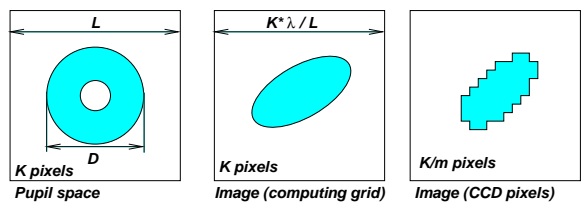}
\caption{Computational grids and scales.
\label{fig:grids}}
\end{figure}


\begin{figure}
\plotone{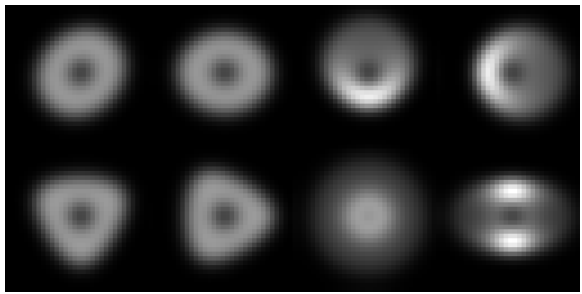}
\caption{Mosaic of 8 defocused  images with Zernike aberrations from 5
  to  12   (left  to  right   and  top  to  bottom)   of  $0.3$~$\mu$m
  amplitude. Seeing $1''$, defocus 3.3~$\mu$m.  Each image is $7.48''$
  wide, 32x32 pixels, $D=4.1$~m.
\label{fig:mosaic}}
\end{figure}

In   our  implementation   of  the   algorithm,  the   computation  of
(\ref{eq:I}) is carried out using  the Fast Fourier Transform (FFT) on
a square numerical grid  of $K\times K$ points (Fig.~\ref{fig:grids}).
The linear size $L$ of the pupil-plane grid should be large enough for
a  telescope   diameter  $D$,  $L   \geq  D$;  critical   sampling  of
diffraction-limited images  requires $L \geq 2D$.  Then  the sampling
in the image space is  $\lambda/L$ (smaller than the diffraction limit
$\lambda/D$) and the  size of the field of view  is $K \lambda/L$.  We
select a  part of the image centered  on the star that  fits into this
field.   In the case  of large  telescopes the  sampling   is fine,
hence we  are forced to  select a large  grid size $K$ to  have enough
field,  at   the  cost  of  slower   calculation.   For  computational
efficiency $K$ has to be an integer power of 2.  The choice $K=256$ is
good for a 4-m telescope.

The CCD  pixels are normally  much larger than $\lambda/D$,  hence the
resulting image  has to be  binned by some  factor $m$. The  number of
``coarse'' CCD pixels is then $N_{CCD} = K/m$. Considering that $K$ is
a power of two, both $m$  and $N_{CCD}$ also have to be integer powers
of two.  Typically, $N_{CCD}=32$ and $m=8$. The CCD pixel size is then
$p = m \lambda/ L$.

The wavefront is represented as a sum of Zernike aberrations up to
some number $N_z$, 

\begin{equation}
W({\bf x}) = \sum_{j=2}^{N_z} a_j Z_j({\bf x}).
\label{eq:W}
\end{equation}
Zernike  polynomials in  the  form of  \citet{Noll}  are used.   Their
amplitudes  (coefficients  $a_j$)  are  equal  to  the  rms  wavefront
variance   over    the   pupil.    The   piston    term   $(j=1)$   is
omitted. Defocused  images (donuts) are obtained by  setting the focus
coefficient $a_4$ to some large positive or negative value.

A  monochromatic  image  computed  from  (\ref{eq:I})  contains  sharp
details of the size  $\lambda/D$ caused by diffraction.  These details
are usually  not seen,  being smoothed by  coarse detector  pixels and
seeing.  In this case  the monochromatic  image model  also represents
broad-band images, and  we can even use a value of  $\lambda$ in the simulation
which is larger than the actual wavelength of observation to, in effect, 
increase the size of the modeled  field.  

The  blur  caused  by  the  time-averaged  seeing  is  modeled  as  a
convolution  with a  Gaussian kernel.   The  FWHM of  the seeing  disk
$\epsilon$  is  proportional   to  the  Gaussian  parameter  $\sigma$,
$\epsilon  =  2 \sqrt  {2  \ln  2}  \sigma \approx  2.35\sigma$.   The
convolution is computed in frequency space by multiplying the FFT
of the image, $\tilde{I}({\bf f})$, by a filter

\begin{equation}
\tilde{I}_s({\bf f})  =   \exp (- 2 \pi^2 \sigma^2 |{\bf f}|^2 )
\label{eq:seeing}
\end{equation}
and doing the inverse FFT. This double FFT is costly in computing time
if  done on  the full  $K  \times K$  grid. When  detector pixels  are
smaller than  $\epsilon$, as  is the case  of astronomical  imagers, a
much  faster calculation  on a  grid  of (binned)  detector pixels  is
justified.  Seeing, together with a set of Zernike coefficients, forms
a vector of parameters that define  the donut model. We put the seeing
in the  first element of this  vector $\epsilon =  a_1$, replacing the
useless piston term.  An example  of donut images corresponding to first few
Zernike aberrations is shown in Fig.~\ref{fig:mosaic}.

\section{Second moments}
\label{sec:mom}

First-order moments  (centroids) of telescopic images  are widely used
for guiding. Here  we show that the second moments  are equally useful for
estimating the second-order aberrations, defocus and astigmatism.

Let $I_{ij}$ be  the image of a point source presented  as an array of
detector pixels  $i,j$.  The coordinates  $x$ and $y$ are  measured in
pixels.  The  zero-order moment $I_0$,  first moments $x_c$  and $y_c$
(in  pixels) and  the second  moments $M_x$,  $M_y$, and  $M_{xy}$ (in
square pixels) are:

\begin{eqnarray}
I_0 & = & \sum I_{ij} \nonumber \\
x_c & = & I_0^{-1} \; \sum x_{ij} I_{ij} \nonumber \\
y_c & = & I_0^{-1} \; \sum y_{ij} I_{ij} \nonumber \\
M_x & = & I_0^{-1} \; \sum (x_{ij}-x_c)^2 I_{ij} \nonumber \\
M_y & = & I_0^{-1} \; \sum (y_{ij}-y_c)^2 I_{ij} \nonumber \\
M_{xy} & = & I_0^{-1} \; \sum (x_{ij}-x_c) (y_{ij}-y_c) I_{ij}  
\label{eq:mom}
\end{eqnarray}

Evident  combinations of  the second moments  relate them  to  defocus and
astigmatism.  Indeed,  the defocus should be proportional  to the size
of the  donut which, in turn,  is the average  of its size in  $x$ and
$y$.  The $45^\circ$ astigmatism  $a_5$ causes image elongation in the
diagonal  direction and  should be  proportional to  $M_{xy}$, whereas
$a_6$ should  be proportional to the  difference of the  image size in
$x$ and  $y$.  Thus, we  introduce the coefficients $A_4$,  $A_5$, and
$A_6$ and  express them  in angular units  (e.g. arcseconds)  with the
help of the angular size of detector pixel $p$:

\begin{eqnarray}
A_4 & = & p \sqrt{(M_x + M_y)/2}    \nonumber \\
A_5 & = & p M_{xy} (M_x M_y)^{-1/4}    \nonumber \\
A_6 & = & 0.5 p (M_x - M_y)  (M_x M_y)^{-1/4}  . 
\label{eq:A}
\end{eqnarray}

\begin{figure}[h]
\plotone{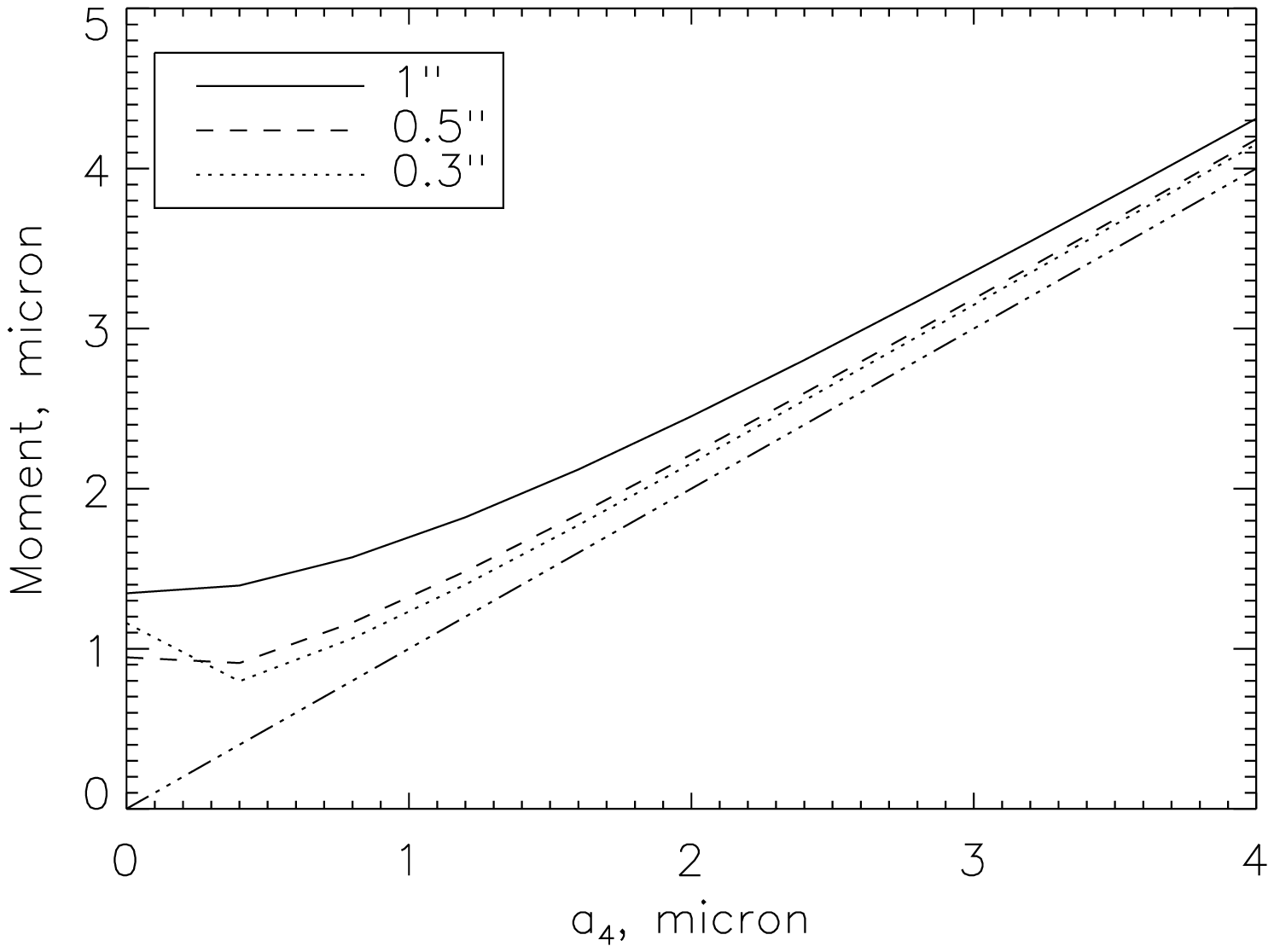}
\plotone{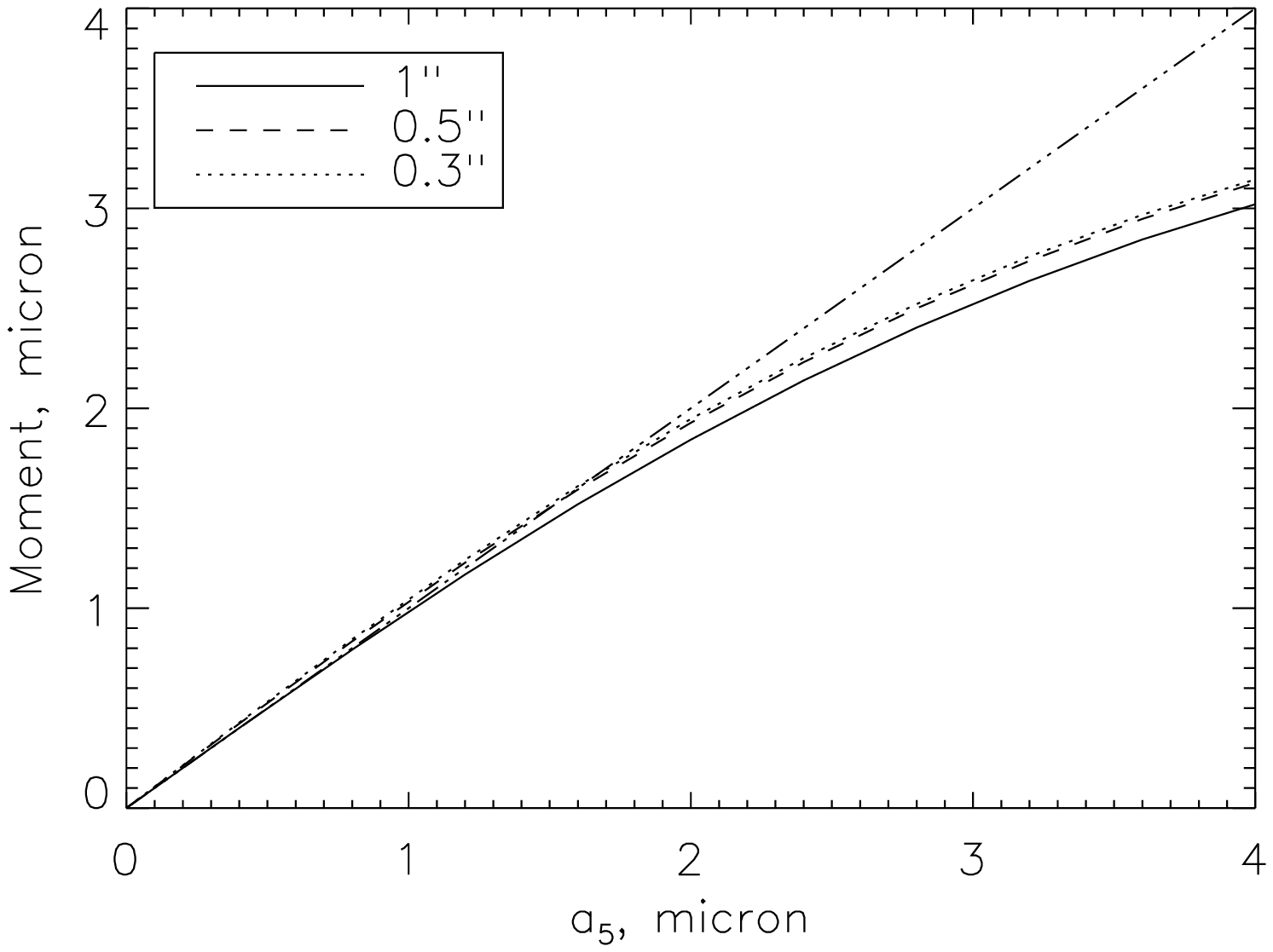}
\caption{Focus aberration $a_4$ (top) and
  astigmatism $a_5$ (bottom) measured by moments, as a function of
  true coefficients. For the astigmatism, the defocus of 3~$\mu$m is
  set. Pixel size 0\farcs5, seeing 0\farcs3, 0\farcs5 and 1\arcsec.  
\label{fig:focus}}
\end{figure}

Next  we must find the relationship between  those coefficients  and the
Zernike   amplitudes.     In the case of defocus,    this   is   relatively
straightforward. The second moment of  a uniform disk of radius $\rho$
is readily calculated to be $M_x = M_y = \rho^2/4$. On the other hand,
the angular radius of the defocused image $\rho$ is found as the first
derivative  of  the  wavefront  at  the  edge of  the  pupil  (in  the
geometrical-optics approximation),

\begin{equation}
\rho = a_4 \frac{8 \sqrt{3}}{D} ,
\label{eq:rho}
\end{equation}
where $a_4$ is the Zernike coefficient of the wavefront.

This leads  to $A_4  = a_4 (4  \sqrt{3})/D$.  There is  similar linear
relation between $A_5$ and $a_5$ with a different coefficient.  We did
not derive this analytically, but  rather found the coefficient by means
of numerical simulation, $A_5 = 0.23 a_5/D$ and $A_6 = 0.23 a_6/D$.

Our  simulations  show that  $A_5$  and  $A_6$  are indeed  very  good
measures  of  the  astigmatism  (Fig.~\ref{fig:focus}). To  the  first
order, they do  not depend on defocus (provided it  is larger than the
astigmatism  itself) and  on other  higher-order aberrations.   On the
other hand,  the linear  relation between $A_4$  and $a_4$  holds only
when the defocus  dominates the seeing blur and  pixel size, and there
is always some bias. 

Second moments  provide an easy and  fast way to  evaluate the defocus
and astigmatism.   To recover the sign of  these aberrations, however,
we need  to know if the  donut is intra- or  extra-focal.  The moments
are  used as  a first  step in  fitting  models to a  donut image.

Second moments are finite in geometrical-optics approximation but they
diverge in physical optics because the intensity of a diffraction spot
does not decrease rapidly enough. Practically, only a finite number of
image pixels is considered, hence the divergence of second moments is
not an issue. 

The computation  of $A_4$  may be  used as a  more efficient  means of
focusing   the  telescope   than  the   traditional   focus  sequence.
Figure~\ref{fig:focus} shows  that a dependence  of the image  size on
the true  focus has  zero slope near  $a_4 =  0$, hence the  method of
focus  sequences (series  of images  near best  focus) has  the lowest
sensitivity to focus and the highest sensitivity to seeing variations.
By  taking one  image sufficiently  far from  focus  and extrapolating
back,  we  obtain  a  better  sensitivity and  less  vulnerability  to
seeing. However, a small bias due to seeing still remains. This can be
eliminated  by taking  two images  with large  defocus  bracketing the
expected true focus.  Let $A_4^+$ and $A_4^-$ be  the focus parameters
(without sign)  derived from these  two images that correspond  to the
focus encoder settings $F^+$ and $F^-$, respectively. Evidently,
\begin{eqnarray}
A_4^+ & = & \alpha (F^+ - F_0) + \delta    \nonumber \\
A_4^- & = &  \alpha (F_0 - F^-) + \delta  , 
\label{eq:foc}
\end{eqnarray}
where $F_0$ is the encoder  setting for perfect focus, $\alpha$ is the
proportionality  coefficient   specific  for  each   telescope,  and
$\delta$ is  the small bias due to  seeing, which we assume  to be the
same  for both  exposures. It  is possible  to determine  two unknowns
$F_0$ and $\delta$ from this system, so the true focus encoder setting
is
\begin{eqnarray}
F_0 = (F^+ + F^-)/2 + (A_4^- - A_4^+)/(2 \alpha) .
\label{eq:F0}
\end{eqnarray}

The reason this method is not in common use at observatories is likely related
to the  need to determine the value of $\alpha$  for each telescope/detector 
 combination and the need to have a reliable focus  encoder. However, the method
should   be   faster   and   more  accurate   than  traditional focus
sequences. Hopefully, it will  become a standard tool in astronomical imaging.

\section{Iterative model fitting}
\label{sec:fit}


\begin{figure}[h]
\epsscale{1.0}
\plotone{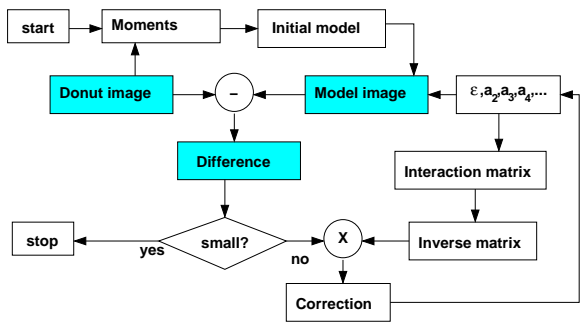}
\caption{Block-diagram of the fitting algorithm.
\label{fig:fit}}
\end{figure}

The  relation between  the phase  aberrations and  resulting  image is
doubly non-linear.  The first  non-linear transformation occurs in the
conversion  from the phase  distribution $\phi$  to the  complex light
amplitude $e^{i  \phi}$.  The second non-linear  transformation is the
calculation of the  intensity distribution as a square  modulus of the
FFT.  Thus,  it is not  possible to fit  a model in  a straightforward
way,  but rather  iterative  methods  have to  be  employed.  At  each
iteration,  small differences  between  the model  and  the image  are
translated into small corrections to the model.

An  insight into  the fitting  process is  provided by  the  theory of
curvature  sensing   \citep{Roddier90}.  A  defocused   image  can  be
considered  as being  an approximate  image  of the  pupil where  each
aberration  produces  a signal  proportional  to  the local  curvature
(Laplacian).  Thus,  in the limit of small  aberrations, the intensity
distribution in the donut can  be represented as the sum of Laplacians
of  the Zernike modes  with suitable  coefficients and  scaling.  This
provides  the required  linearization for  deriving the  correction at
each iteration  step. In other words,  a combination of  a large known
aberration  (defocus) with  small high-order  aberrations leads  to an
approximate linearization of  the image-formation process with respect
to high-order terms.

The method  of modeling the donut is  as follows (Fig.~\ref{fig:fit}).
The first estimate of the  Zernike coefficients up to $a_6$ is derived
by the method of moments (we  initially set $a_1 = 0\farcs5$).  At the
second step,  the gradients of the  model with respect to  each of the
parameters  are computed as  differences between  the model  image and
images obtained by varying each Zernike coefficient by a small amount.
These  differences  are computed  for  each  pixel  of the  image  and
combined in the  {\it interaction matrix} $H$ of  the size $N_p \times
N_z$, where $N_p$ is the total number of pixels in the image and $N_z$
is  the number  of fitted  Zernike terms.   This matrix  relates small
variations of the parameters  (seeing and Zernike coefficients) to the
variations of the signal --  intensities in each pixel.  The seeing is
considered as an additional  unknown parameter and fitted jointly with
the aberration coefficients.

The matrix $H$  is inverted, so the differences  between the model and
the actual image can be  converted into the corrections to the Zernike
coefficients.  The  new set  of coefficients is  the new  model which,
hopefully, is  a better  approximation of the  donut.  The  process of
image formation  being non-linear, we  have to repeat  this linearized
fitting again  and again iteratively  until the model  converges.  The
algorithm is  similar to  the closed-loop wavefront  control algorithm
used  in  adaptive  optics:  at  each iteration  we  obtain  a  better
approximation to the donut. Further details are given in the Appendix.

The number of ``resolution elements'' across the pupil is of the order
$2  \rho /\epsilon$.   Thus, if  aberrations of  high-order are  to be
measured, a larger donut radius  $\rho$ is needed.  On the other hand,
curvature  sensors are  known to  suffer from  severe  aliasing, where
un-modeled    high-order    aberrations    distort    the    low-order
coefficients.  Hence,  a  defocus  of  $2  \rho/\epsilon  \sim  n$  is
recommended for sensing aberrations up to the radial order $n$.  These
considerations are further developed in the next Section.

\section{Performance of the donut algorithm}
\label{sec:perf}

\subsection{Aliasing}

\begin{figure}[hb]
\plotone{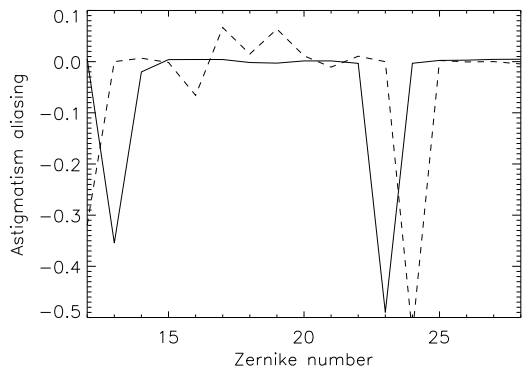}
\caption{Aliasing  coefficients of  Zernike  astigmatisms $a_5$  (full
  line)  and $a_6$  (dash).   Seeing $1''$,  pixel scale  $0\farcs23$,
  defocus $2 \rho = 3.1''$, modeling up to $N_z = 11$.
\label{fig:alias}}
\end{figure}

Suppose we  want to measure  Zernike coefficients up to  11 (spherical
aberration) by fitting a model to the donut. To what extent is the result
distorted by the presence of  high-order aberrations?  Let $a_k \neq 0 $
be the amplitude of un-modeled high-order aberration ($k > N_z$) which
produces an error  $\Delta a_j$ of the $j$-th  coefficient.  The ratio
$\Delta   a_j   /a_k$    is   called the  {\it   aliasing   coefficient}.
Figure~\ref{fig:alias}   plots  these  coefficients   for  astigmatism
($j=5,6$).   The  $a_5$  term  is  aliased mostly  with  $a_{13}$  and
$a_{23}$ assuming seeing of $1''$.  The condition  $2  \rho/\epsilon \sim  n$ is  approximately
satisfied in this example. However, if the seeing improves to $0.5''$,
the aliasing coefficient with $a_{13}$ increases from $-0.35$ to $+2$.

Clearly, aliasing  can be a problem for  a donut sensor, as  it is for
any curvature  sensor.  The evident  solution, though, is  to increase
the order of the fitted model until all aliased terms are explicitly taken
care of. Another way to reduce the aliasing is to decrease the defocus
to  the  minimum  value  required   to  measure  a  selected  set  of
low-order aberrations.

For comparison, we  studied the aliasing of astigmatism  measured by a
2x2 S-H sensor.  We find that, if the full telescope aperture is used,
the aliasing  coefficient of $a_5$  with $a_{13}$ is $+1.4$,  and that
the aliasing  coefficient is even  larger for some higher  terms.  The
aliasing of  an S-H sensor can  be reduced by reducing  the portion of
the aperture used  for a 2x2 sensor or by increasing  the order of the
sensor.  It is clear, however, that aliasing in a low-order S-H sensor
is  of the  same order  as  for the  donut method,  with less  options
available for decreasing it.

\subsection{Detector noise}


\begin{figure}
\plotone{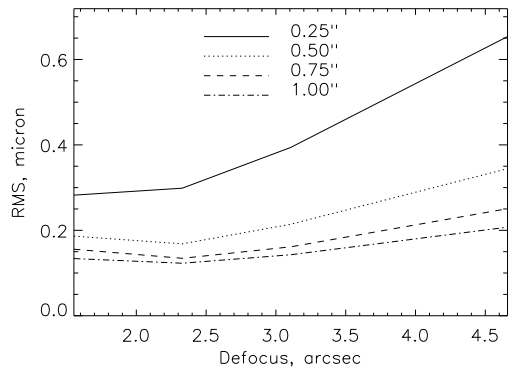}
\caption{The  rms  noise  of  the astigmatism  coefficient  $a_5$  for
  various  diameters  of the  donut  and  different  CCD pixel  scales
  (indicated  on  the  plot)  under  $1''$ seeing.   Readout  noise  10
  electrons, $N_{ph}= 1000$.
\label{fig:ast2}}
\end{figure}


\begin{figure}
\plotone{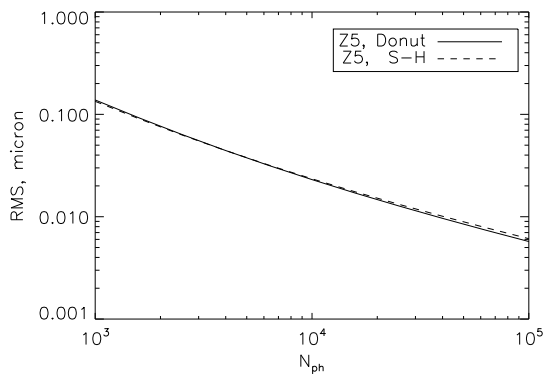}
\caption{The rms noise of  the astigmatism coefficient $a_5$ vs. total
  number of photons $N_{ph}$ for donut method ($2 \rho =  2\farcs5$, pixel
  size  $p=0\farcs75$, readout  noise $R=10$)  and  for a  2x2 S-H  sensor
  ($p =0\farcs75$, $R=10$). In both cases seeing is 1\arcsec.
\label{fig:noisecomp}}
\end{figure}

In some instances it is  important to measure optical aberrations with
relatively faint stars.  The readout  and photon noise may then become
an issue because the light in  a donut is spread over many CCD pixels.
The problem can be studied by  simulating a series of noisy images and
evaluating the scatter of  the measured Zernike coefficients. However,
a much faster analytical evaluation of the errors is available through
the  covariance matrix,  Eq.~\ref{eq:Cz}. We  have verified  that this
method gives an answer which  is consistent with the results of direct
Monte-Carlo simulation.

For  a given total  number of  detected photons  $N_{ph}$ and  a given
readout noise $R$, the  errors of measured Zernike coefficients depend
on  the size  of the  donut, the  size of  detector pixel,  seeing and
aberrations. In  the following we  assume that all  aberrations except
defocus  are corrected, as  would be  appropriate in  an active-optics
application; if this is not true, the results would be different.

An example of  optimization for measuring $a_5$ under  $1''$ seeing is
shown  in Fig.~\ref{fig:ast2}  for a  faint  star, when  the noise  is
mostly  dominated by the  detector readout  noise.  The  optimum pixel
scale  in  these conditions  is  about  $1''$  and the  optimum  donut
diameter  is  about $2.5''$.   However,  large  deviations from  these
optimum values cause only a  minor increase of the noise.  The optimum
parameters depend  on the  Zernike number, on  seeing and on  the flux
$N_{ph}$. A  reasonable choice of parameters  can be made  to ensure a
near-optimum measurement  of several Zernike coefficients  for a range
of seeing conditions.

In the case of faint  stars when  the detector  noise $R$  dominates, the
errors of the Zernike coefficients must be proportional to $R/N_{ph}$. The
calculations show  this to  be approximately true  up to  $N_{ph} \sim
10\;000$ (for our choice of $R=10$). At larger flux, the errors improve
only as  $1/\sqrt{N_{ph} }$.  However, the photon-noise  errors in the
bright-star regime  are so small that they  become irrelevant compared
to other errors.

The intensity modulation in  the donut increases with increasing number
$j$ (at constant amplitude  $a_j$), because it is roughly proportional
to the curvature.  Equating the modulation with noise,  we expect that
noise propagation decreases  with $j$.  This is indeed  the case.  One
notable exception, however, is the spherical aberration which can have
an error much larger than other terms of the same order. We trace this
anomalous  behavior to  the  cross-talk between  $a_{11}$ and  seeing.
Indeed, processing of  real data shows that the  estimates of $a_{11}$
and $\epsilon$ are often less repeatable, compared to other terms.

We  compared  the  sensitivity  of  the  donut  method  for  measuring
astigmatism  with  that of  a  2x2 S-H  sensor  and  found that  their
performance   in   the   low-flux   regime   can   be   very   similar
(Fig.~\ref{fig:noisecomp}).   The  noise  was computed by the same 
method for both measurement techniques i.e.   by relating errors of
pixel  intensities directly  to  the errors  of Zernike  coefficients.
This should give the  lowest possible error.  In practice, aberrations
are  normally derived in  a S-H  sensor from  centroids of  the spots,
hence with  somewhat larger errors.   Naturally, the noise  depends on
the parameters such as defocus,  seeing, and pixel size, hence in some
situations S-H sensors can  perform better than donut. S-H  is to be preferred
for   measurement  of   atmospheric  tip-tilt   errors.    The  formal
sensitivity of  donut to  tip and tilt  is only slightly  inferior
to that of S-H, but at short exposures the centroids of the donut
images will be severely displaced by higher-order aberrations and will
not provide good measurements of tilts.

\subsection{Convergence and reliability}

The  iterative fitting has  been tested  on different  simulated donut
images and  always produced  the expected result.  However, processing
real  data  is sometimes  more  tricky.   The  interaction matrix  $H$
depends on  the aberrations, it  changes between different  images and
even during the  fitting of one image. When a  large number of Zernike
terms is considered, it is  common to encounter low singular values in
$H$.  This  means that  some combinations of  parameters are  not well
constrained by  the data, hence  the noise will be  amplified. Leaving
such combinations  un-fitted by  rejecting small singular  values does
not solve the problem because we  may obtain a good model of the donut
image  with a  parameter set  which is  very different  from  the true
parameters. This  happens when significant  high-order aberrations are
present,  but the  defocus is  not large  enough, i.e.  in  the caustic
regime.

One  way  to  get  around  this problem  is  to  determine  high-order
aberrations  separately (e.g. by  fitting a  bright-star image  with a
large defocus)  and then  to include them  in the model  for low-order
fits.   Including  such  pre-defined  parameters (we call them  static
aberrations) improves  the convergence and the  fit quality. Low-order
fits  are more  stable and  give reproducible  results.   However, the
coefficients of  low-order aberrations derived  in this way  depend on
the adopted  static aberrations: a  different result is  obtained from
the  same  data when  a  different  vector  of static  aberrations  is
supplied initially.

\subsection{Other error sources}

In real life, optical aberrations in the beam change with time because
of the instability of  telescope optics, the changing refractive index
of the  air in the  dome, and seeing.   Averaging donut images  over a
sufficiently long time $T$ (typically 10-30s) reduces the contribution
of  variable aberrations only  by a  factor of  $\sqrt{\tau/T}$, where
$\tau$ is the time constant  of the variation.  Consider, for example,
a 4-m telescope  with 5~m/s wind and $1''$  seeing.  The rms amplitude
of the random astigmatism caused by the seeing is 270~nm, according to
the formulae of \citet{Noll}, and  its time constant is 0.25~s.  Thus,
in a  10-s exposure  we expect  a random error  of astigmatism  of the
order of 40~nm, or larger if  the wind is slow and/or some aberrations
are produced  by air  inside the dome.   The statistical noise  can be
reduced by taking  longer exposures but may still  remain a dominating
error source.

If the donut image is blurred in one direction by imperfect guiding or
telescope  shake during the  exposure, this  departure from  the ideal
model will result in spurious aberrations, mostly astigmatisms of 2-nd
and 4-th order.  Simulations for the  case of SOAR show that a blur of
$1''$ causes  errors of  $a_6$ and $a_{12}$  of only 20~nm,  a smaller
blur  has negligible effect.   Hence the  blur is  never a  problem at
modern telescopes with good tracking.

\section{Examples}
\label{sec:exa}

\subsection{Internal consistency}

\begin{figure}[ht]
\plotone{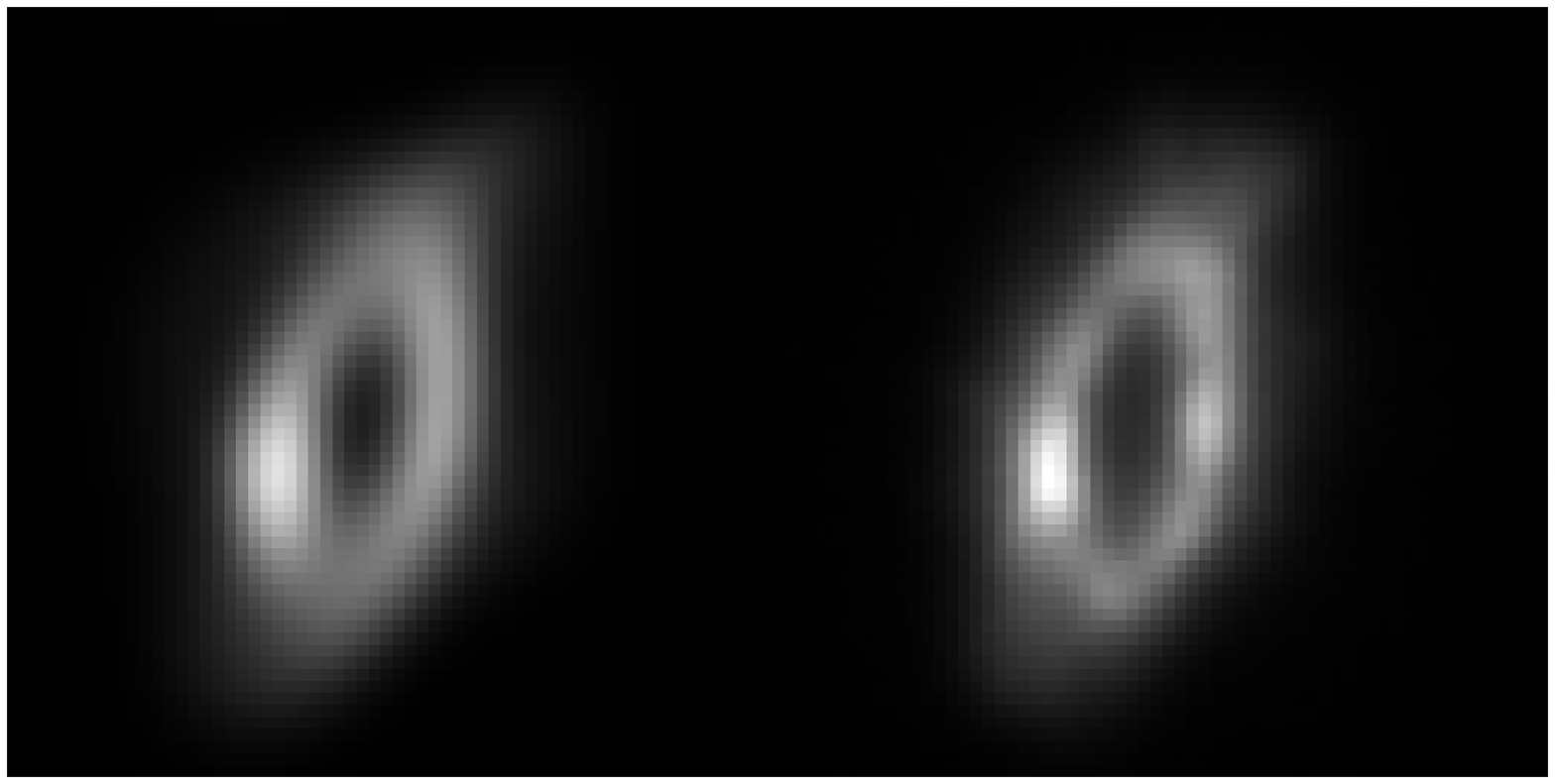}
\plotone{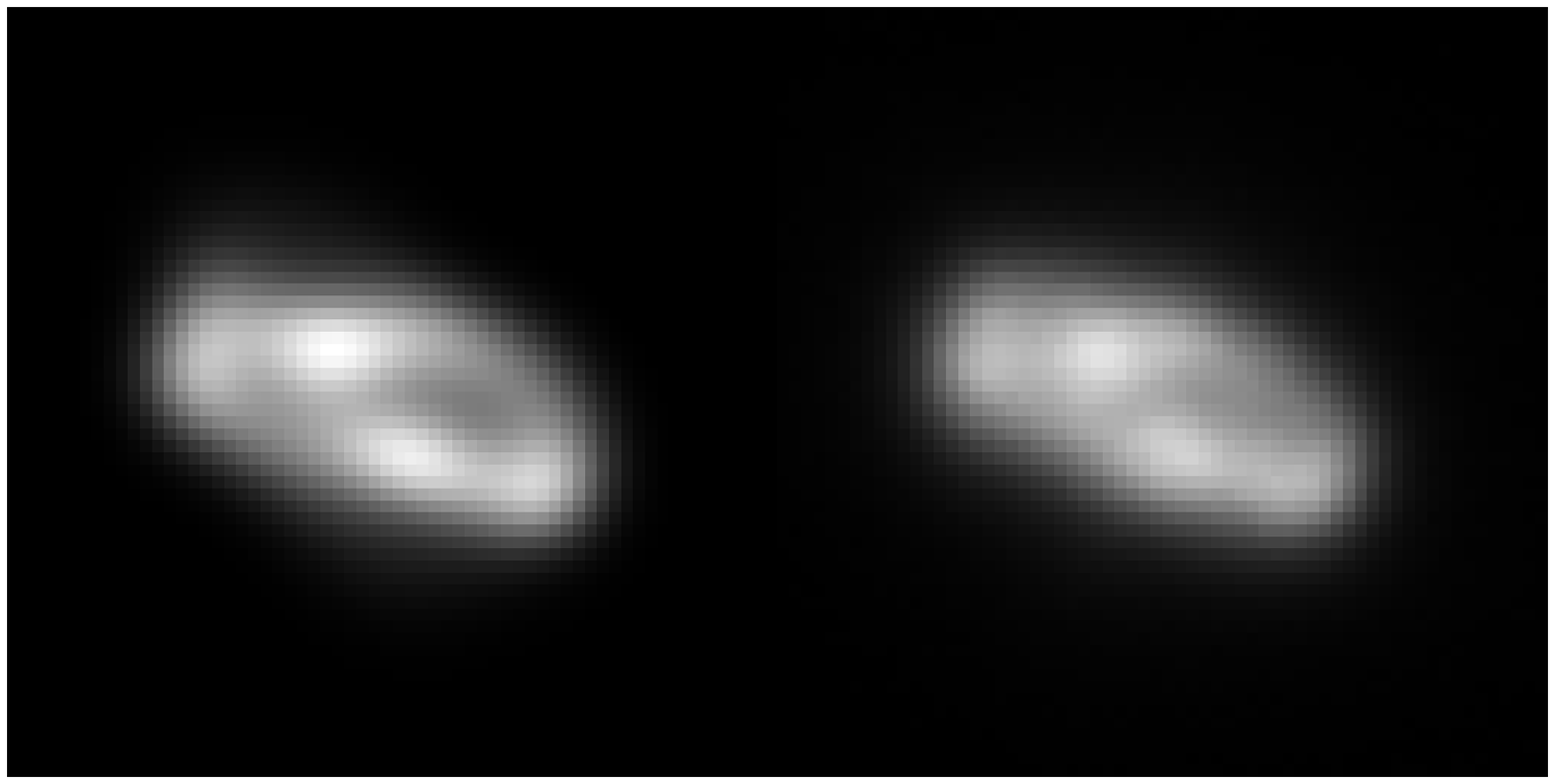}
\caption{Intra-focal (top) and  extra-focal (bottom) astigmatic images
taken at SOAR on March 6/7 2005 (on the right) and their corresponding
models (on the left). Pixel  size 0\farcs154, field of view 9\farcs85.
The exposure numbers are 113 (top) and 115 (bottom).
\label{fig:113}}
\end{figure}



\begin{table*}
\caption{Some Zernike coefficients ($\mu$m rms) measured on SOAR images
  with artificial astigmatism.}
\label{tab:result}
\begin{tabular}{l ccc ccc}
\hline
Image & Seeing, \arcsec & $a_4$ & $a_5$ &   $a_6$ &  $a_7$ &  $a_8$ \\
\hline
113 &   0.936 &  $-$3.704 & $-$1.061 &  1.205 &  0.042 &  0.126  \\
114 &   0.978 &  $-$3.537 & $-$1.165 &  1.264 & $-$0.006 &  0.130  \\
115 &   1.211 &   3.271 & $-$1.225 &  1.239 & $-$0.055 &  0.033 \\
116 &   1.090 &   3.028 & $-$1.487 &  0.852 &  0.077 & $-$0.242 \\
137a  &   0.871 & $-$4.668 & $-$1.446 &  0.133 &  0.570 &  0.783 \\  
137b  &   0.851 & $-$4.590 & $-$1.431 &  0.135 &  0.555 &  0.762 \\ 
137c  &   0.858 & $-$4.645 & $-$1.426 &  0.080 &  0.623 &  0.783 \\ 
137d  &   0.884 & $-$4.853 & $-$1.504 &  0.185 &  0.630 &  0.770 \\ 
\hline
\end{tabular}
\end{table*}

Several series of defocused images were taken at the SOAR telescope in
March 2005 and processed with  the donut algorithm.  One example shown
in Fig.~\ref{fig:113} was acquired with a pixel scale of $0\farcs154$ and
25-s exposure  time using a conveniently bright  star.  An astigmatism
was  introduced  intentionally by  de-tuning  the actively  controlled
primary   mirror.    Extra-  and   intra-focal   images  were   fitted
independently  of each other  with $N_z  = 28$  terms.  At  each focus
setting, two  images were acquired.  The defocus  of 3~$\mu$m produces
donut     images     of     4\farcs2    diameter.      The     results
(Table~\ref{tab:result})  show a good  coherence of  the measurements,
irrespective of which side of the focus they were taken. The residuals
between  model  and  image are  from  5\%  to  9\%.  The  presence  of
uncorrected (but well-modeled)  high-order aberrations is evident in
Fig.~\ref{fig:113}.

Yet another  test was  done by fitting  defocused images  of different
stars in  the same exposure.  The flux in  the image 137a is  30 times
more than in the image 137d, yet the Zernike coefficients derived from
these images  agree well  (Table~\ref{tab:result}). Here, the  fit has
been limited  to 11  terms (with static  aberrations up  to $a_{28}$),
because full  fitting of 28  terms did not give  reproducible results.
This  instability  is  apparently  caused  by  significant  high-order
aberrations, as seen in Fig.~\ref{fig:113}.

An estimate of the internal  accuracy of the donut method was obtained
by  processing several  consecutive  images. The  rms  scatter of  the
coefficients for 2-nd and 3-rd order aberrations ranges typically from
0.05 to 0.15~$\mu$m for 60-s exposures.

\subsection{Comparison with a  Shack-Hartmann WFS}

\begin{figure}[ht]
\plotone{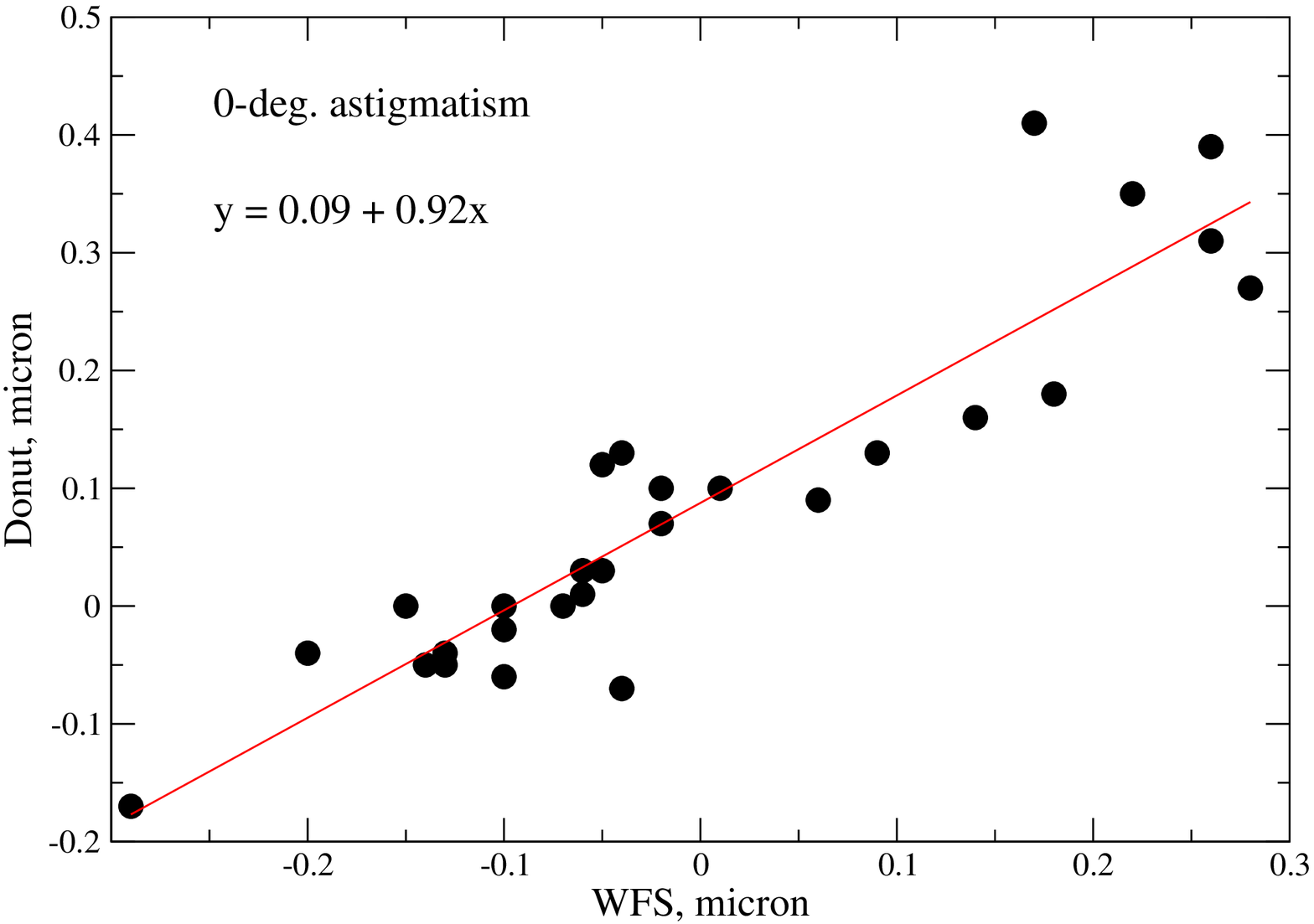}
\plotone{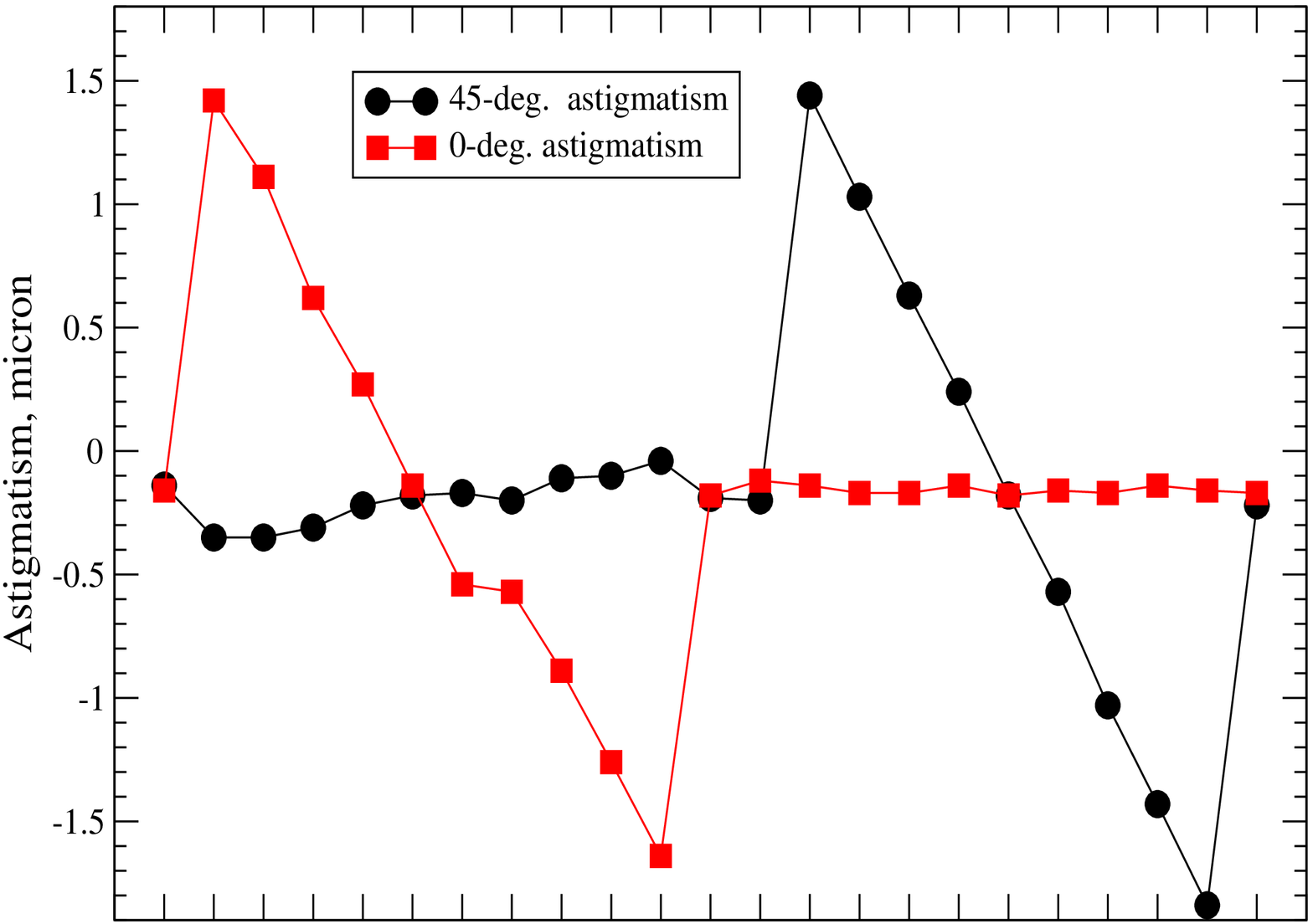}
\caption{Comparison between  donut and CWFS at  SOAR.  (a) Astigmatism
 changes caused  by the telescope  motion in elevation as  measured by
 the CWFS  (horizontal axis) and  donut (vertical axis). The  data was
 taken on April 13/14 2006.  (b) Two astigmatism coefficients measured
 with  donut as  the mirror  shape is  de-tuned with  an  amplitude of
 $\pm$1~$\mu$m and step 0.25~$\mu$m (April 15/16, 2006).
\label{fig:CWFS}}
\end{figure}


\begin{figure}
\plotone{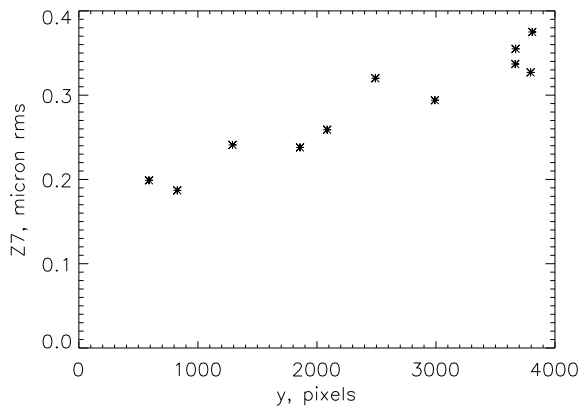}
\caption{Variation of the coma coefficient $a_7$ across the field in one of
  the detectors of the Mosaic imager on the Blanco telescope. 
\label{fig:coma}}
\end{figure}

The  donut   method  has  been  compared  with   the  SOAR  high-order
Shack-Hartmann control  WFS (CWFS) that  is part of  the active-optics
system  used for  tuning  the  primary mirror.   The  response of  the
primary  mirror   actuators  was  calibrated   independently  by  the
manufacturer  and is  $\sim  1.6$ times  larger  than the  aberrations
measured by the CWFS.

 The donut  data were  taken with  the SOAR imager  and binned  to the
pixel scale  of $0\farcs154$.  Three  60-s exposures for  each setting
were processed independently, providing an estimate of the measurement
errors.   The CWFS data  are single  measurements with  10~s exposure,
more vulnerable  to the  insufficiently averaged atmospheric  and dome
turbulence  than donuts.   The measurements  with donut  and  CFWS are
sequential  as  these devices  occupy  different  focii  of SOAR.  The
Zernike  coefficients obtained  with donut  were rotated  to  the CFWS
system by the known angle between these instruments.  Both instruments
give  Zernike  coefficients  on  the  same scale  --  rms  microns  of
wavefront aberration.

Figure~\ref{fig:CWFS}a shows a comparison between the two sensors
as  the telescope  was  tipped  in elevation  and  brought back.   The
systematic  trend  of  the  $0^\circ$ astigmatism  with  elevation  is
evidenced by both methods, with  some offset and scale factor apparent
from  the  linear  regression.  The  scatter  of  points  around  this
regression is typical for such  tests and compatible with the internal
consistency of each method.

For  another test,  the shape  of the  SOAR primary  was  distorted by
``dialing  in'' astigmatism coefficients  in $0^\circ$  and $45^\circ$
with a  full amplitude  $\pm 1$~$\mu$m and  a step  0.25~$\mu$m (these
numbers refer to  the primary mirror aberrations as  determined by the
manufacturer).  The mirror was  initially flattened with the CFWS. The
result  (Fig.~\ref{fig:CWFS}b) shows  that the  donut  method measures
these aberrations with a coefficient  of $\sim 1.6$ (same as the CWFS)
and an offset presumably arising  from the fixed difference of optical
aberrations between the focii of CWFS and imager.

\subsection{Mosaic imager at the Blanco telescope}

The classical  4-m Blanco telescope  at Cerro Tololo is  equipped with
the  wide-field CCD mosaic  at its  prime focus.   The pixel  scale is
$0\farcs27$.  We  processed donut  images extracted from  the standard
focusing  sequences (exposure  time  10~s per  focus setting,  maximum
defocus 1.5 to  2~$\mu$m).  Although these data were  not intended for
the  aberration analysis,  fitting them  with donut  models  was quite
successful,  with a  typical rms  intensity  residuals of  6\% for  28
Zernike terms.  The fitting takes  20--30~s on a 1~GHz PC with $K=256$
grid.

Comparing  the coefficients of  low-order aberrations  determined from
the first  and the  last images  in each sequence,  we find  a typical
difference  of 0.1~$\mu$m  or  less,  i.e. similar  to  the SOAR  data
presented in  Table~\ref{tab:result}. The most likely  reason of these
differences  is  a  real  slow  variation of  the  aberrations  between
exposures in the focusing sequence.

We processed the  first image of the focusing  sequence extracted from
11 different stars  in one of the detector  segments. These images are
simultaneous and the scatter of the measured coefficients in this test
was much smaller, from 0.025  to 0.073~$\mu$m. Part of this scatter is
caused   by   real   variations   of  the   aberrations   across   the
field.  Figure~\ref{fig:coma}  shows  a   clear  trend  in  the  coma
coefficient $a_7$ as a function of the $y$-coordinate of the star.

This example shows how  a quantitative analysis of optical aberrations
can be done with simplicity, as a by-product of standard observations.
It is possible to measure aberrations across the field of a prime-focus
camera with a  Hartmann mask, but the donut  technique makes this task
much easier. The rms accuracy can reach 25~nm, or $\lambda/25$.


\section{Conclusions}
\label{sec:concl}

We   have  shown  that   focus  and   astigmatism  can   be  evaluated
quantitatively from the second moments of defocused images. One useful
application  of this  analysis will  be a  fast and  accurate focusing
procedure for  classical imaging, suggested  here as a  replacement of
traditional  focusing  sequences.  Furthermore,  donut  images can  be
fitted directly to a set of Zernike coefficients (complemented with an
additional  parameter, seeing),  offering a  practical way  to measure
aberrations and to tune the optics of ground-based telescopes.

The  donut  method  proposed  here  is  different  from  the  standard
curvature sensing in several aspects.  First, only one defocused image
is needed.   Second, no simplifying  assumption of linearity  is made,
hence the  defocus may be  quite small while measuring  aberrations of
significant amplitude -- comparable to the defocus itself. Third, we do
not use the intensity  transport equation \citep{Roddier90} but rather
compute  the image  model by  a full  Fraunhofer  diffraction integral
using an FFT.  Finite  detector pixel size and additional  blur caused by
the  seeing are  explicitly  taken into  account.   These two  effects
usually  wash  out  any  traces  of  diffraction,  so  the  calculated
monochromatic image is a good model of a wide-band image as well.

The  down-side  of the  full  diffraction  image  modeling is  a  slow
calculation time (a few  seconds for a 4-m telescope)  and a restriction
of the  modeled field of  view.  The donut  method will work  best for
small defocus  values and for measuring low-order  aberrations. On the
other  hand, classical curvature  sensing would  be probably  a better
choice  for high-resolution  sensing, where  a wave-front  map (rather
than Zernike coefficients) is sought.

We plan to apply the donut technique to the closed-loop control of the
SOAR  active  optics  and   to  optical  characterization  of  other
telescopes at CTIO. The method seems to be simple and mature enough to
be offered to other interested users. So far, it is implemented in the
IDL language.

\acknowledgments

We  thank  D.~Maturana,   S.~Pizarro  and  H.E.M. Schwarz  for  taking
defocused images at SOAR, B.~Gregory for processing the images and his
valuable comments, A.~Rest for the help in extracting the Mosaic data.
The  comments of P.~Hickson  on the  draft version  of this  paper are
gratefully acknowledged.

\appendix
\section{Fitting algorithm}
\label{App}

The interaction  matrix $H$ relates intensity changes  in the detector
pixels to the variation of  the Zernike coefficients.  The size of $H$
is  $N_p \times  N_z$ elements,  where $N_p$  is the  total  number of
pixels in the  modeled donut image and $N_z$ is  the number of modeled
Zernike terms (including seeing  which replaces the piston term).  The
pixels are  re-indexed sequentially, $i=1,2,...,\;  N_p$.  The initial
vector  of parameters  ${\bf a}  =  \{ a_1,a_2,...,\;  a_{N_z} \}$  is
supplied at  the beginning, with the first  element $a_1$ representing
seeing. Our task is to find the estimate of ${\bf a}$ that ensures the
best correspondence between the model  $M({\bf a})$ and the image $I$.
Both  model  and  image  are  normalized  to keep  the  sum  of  pixel
intensities equal to one.

We compute $H$  by varying each Zernike coefficient  by a small amount
$\Delta a_j = 0.5/n$ radians, $n$ being the radial order. This choice of
decreasing  amplitudes ensures that the image variations remain in  the linear
regime. The seeing is changed by $\Delta a_1 = 0\farcs1$. So, a $j$-th
column  of $H$  is equal  to the  normalized difference  between pixel
values of  the image  model $M_i$ that  result from changing  the $j$-th
term,
\begin{equation}
H_{i,j} = [M_i({\bf a} + \Delta a_j) - M_i({\bf a})] /  \Delta a_j  .
\label{eq:Hcalc}
\end{equation}
A  large economy  of calculation  is  achieved by  saving the  complex
amplitude  at  the  telescope  pupil  for a  given  model  ${\bf a}$.   When
re-calculating the image with just one modified Zernike term $a_j$, we
only need to multiply the  saved amplitude by the  factor
$\exp [ 2 \pi i \Delta a_j Z_j( {\bf x})/\lambda ]$, instead of
re-computing all Zernike terms.

The interaction matrix  $H$ is inverted in the  sense of least-squares
(LS),
\begin{equation}
H^* = (H H^T)^{-1} H^T .
\label{eq:H*}
\end{equation}
The inversion  of $ (H H^T)$  is done by  Singular Value Decomposition
\citep{NumRec}, rejecting  weak singular values  below some threshold.
This   guarantees  that  poorly   measurable  combinations   of  model
parameters do not  lead to increased noise. In  fact, we progressively
decrease the threshold during iterations when they converge (i.e. when
the residuals decrease), but reset it to the original high value (0.05
of the maximum singular value) in the case of divergence.

The  inverse matrix  $H^*$ is  multiplied by  the vector  of intensity
differences between the input image $I$ and the model image $M$ to get
the correction of the Zernike coefficients $\Delta {\bf a}$:

\begin{equation}
\Delta {\bf a} = H^* \times (I-M) .
\label{eq:da}
\end{equation}

This  equation,  however, treats  all  pixels  with  equal weight.   A
somewhat more  rigorous approach takes into account  the detector and
photon noise,  which differs from pixel to pixel.  Let  the rms  detector noise
(RON) be  $R$ and the total number  of photons in the  image (flux) be
$N_{ph}$.  The  pixel intensities $I_{i}$ are normalized
so  that  $\sum   I_{i}  =1$.   In  this  case   the  noise  variance
$\sigma^2_{i}$ of the measured intensity in a pixel $i$ is

\begin{equation}
\sigma^2_{i} = N_{ph} I_{i} + R^2 .
\label{eq:sigma2}
\end{equation}

A flavor of  LS fitting to data with  variable and un-correlated noise
is  known as  {\it weighted  least-squares}.   If the  columns of  the
interaction matrix $H$  and the residuals $(I-M)$ are  both divided by
$\sigma_{i}$, the problem is reduced  to the LS fitting with constant
noise.   The  weighted  interaction   matrix  $H'$  replaces $H$ in
(\ref{eq:H*}) to calculate $H^*$.  

The   data  vector   $(I - M)$   normalized  by   $\sigma_{i}$  has
uncorrelated elements with unit  variance. Hence the covariance matrix
of  the  restored  Zernike  coefficients  $C_a$  is  simply  related  to  the
restoration operator $H^*$,

\begin{equation}
C_a = \langle \Delta a_j \; \Delta  a_k \rangle =  H^* (H^*)^T.
\label{eq:Cz}
\end{equation}
The variances of the measured Zernike coefficients caused by noise are
equal to the  diagonal elements of $C_a$. This  provides an evaluation
of the noise component of measured aberrations related to the detector
and the brightness  of the star.  In practice, however,  we do not use
the normalization by $\sigma_i$ because it gives high weight to pixels
outside donut and often prevents the convergence.

The quality of the fit is characterized by the relative residuals $Q$,
\begin{equation}
Q^2 = \sum_i (I_i - M_i)^2 /  \sum_i I_i^2.
\label{eq:Q}
\end{equation}
The  iterations continue  until a  condition  $0.99 \;  Q_{\rm old}  <
Q_{\rm new} < Q_{\rm old}$ is reached, i.e.  when the residuals do not
decrease  significantly  any  more.   Reasonably good  fits    have
$Q<0.1$.  To  ensure robust  convergence, we add  at each step  only a
fraction of the computed correction,  $0.7 \Delta {\bf a}$.  If at the
next iteration $Q$  increases, the parameters are not  changed at all,
but  the SVD threshold  is re-set  to a  high value.   The interaction
matrix is re-computed only at even iteration steps, to save time.





\end{document}